\documentclass[useAMS,usenatbib,usegraphicx]{mn2e}

\title[Regenerated emission from GRB afterglows]{Does regenerated
emission change the high-energy signal from gamma-ray burst afterglows?}
\author[S. Ando]{Shin'ichiro Ando\thanks{E-mail:
ando@utap.phys.s.u-tokyo.ac.jp}\\Department of Physics, School of
Science, The University of Tokyo, 7-3-1 Hongo, Bunkyo-ku, Tokyo
113-0033, Japan}
\begin{document}

\date{Submitted 9 June 2004; accepted 8 July 2004}

\pagerange{\pageref{firstpage}--\pageref{lastpage}} \pubyear{2004}

\maketitle

\label{firstpage}

\begin{abstract}
We study regenerated high-energy emission from the gamma-ray burst (GRB)
 afterglows, and compare its flux with the direct component from the
 same afterglow.
When the intrinsic emission spectrum extends to TeV region, these very
 high-energy photons are significantly absorbed by the cosmic infrared
 background (CIB) radiation field, creating electron/positron pairs;
 since these pairs are highly energetic, they can scatter the cosmic
 microwave background radiation up to GeV energies, which may change the
 intrinsic afterglow light curve in the GeV region.
Using the theoretical modeling given in literature and the reasonable
 choice of relevant parameters, we calculate the expected light curve
 due to the regeneration mechanism.
As the result, we find that the regenerated emission could only slightly
 change the original light curve, even if we take a rather large value
 for the CIB density, independently of the density profile of
 surrounding medium, i.e., constant or wind-like profile.
This ensures us the reliable estimation of the intrinsic GRB parameters
 when the high-energy observation is accessible, regardless of a large
 amount of uncertainty concerning the CIB density as well as
 extragalactic magnetic field strength.
\end{abstract}


\section{Introduction}
\label{sec:Introduction}

Gamma-ray bursts (GRBs) are known to be highly energetic astrophysical
objects located at cosmological distance.
Accumulated data of many GRBs strongly support relativistic fireball
scenario, in which $\gamma$-rays up to MeV are attributed to internal
shocks due to collisions between fireball shells, while their transient
component, afterglow, from radio to X-rays is attributed to external
shocks due to the interaction of the fireball with external medium.
In addition to such signals, very high-energy photons that range from a
few tens of MeV to GeV have been detected \citep{Hurley94}, and further
the detection of an excess of TeV photons from GRB 970417a has been
claimed with a chance probability $\sim 1.5\times 10^{-3}$
\citep{Atkins00}.
Although the statistics of these high-energy signals are not sufficient
yet, planned future satellites or detectors will promisingly enable us
to discuss high-energy emission mechanisms of GRBs.

Several emission mechanisms of GeV--TeV photons are proposed, such as
synchrotron self-inverse Compton (IC) emission of the electrons
\citep*{Meszaros94,Waxman97,Panaitescu98,Wei98,Wei00,Dermer00a,
Dermer00b,Panaitescu00,Sari01,Zhang01} and the proton-synchrotron
emission \citep{Vietri97,Bottcher98,Totani98b}, as well as some other
hadron-related emission components \citep{Bottcher98}.
These mechanisms could be valid for internal shocks, external forward
shocks, or external reverse shocks of GRBs.

Regardless of the emission mechanism, high-energy photons above $\sim
100$ GeV are expected to be attenuated via the $\gamma\gamma\to e^+e^-$
process.
Target photons with which the initial high-energy photons interact are
the GRB emission itself or the cosmic infrared background (CIB).
As for the latter, it is suggested that such very high-energy photons
may largely be absorbed during its propagation, if the GRB location is
sufficiently cosmological as $z\ga 1$
\citep*{Stecker92,MacMinn96,Madau96b,Malkan98,Salamon98}.
Therefore, the detection of TeV photons from cosmological GRBs will be
very difficult.
However, since created electron/positron pairs due to the interaction
with CIB photons are very energetic, they can IC scatter on the most
numerous cosmic microwave background (CMB) photons, giving rise to a
delayed secondary MeV--GeV emission (\citealt{Plaga95};
\citealt{Cheng96}; \citealt{Dai02a}; \citealt{Wang04};
\citealt*{Razzaque04}).
These regenerated emissions would, therefore, be indirect evidence of
the intrinsic TeV emission as well as a probe of the CIB radiation
field, which is not satisfactorily constrained.
It has been argued that the internal shocks \citep{Dai02a,Razzaque04} as
well as the prompt phase of the external shocks \citep{Wang04} are
possibly responsible for these delayed MeV--GeV emission, and it would
be distinguishable from a different delayed GeV component due to the
direct IC emission from the afterglow.

Future detection of the direct MeV--GeV emission predicted from the
afterglow, would be a probe of an emission mechanism of the high-energy
region as well as physical parameters of the fireball.
However, the primary emission is possibly modified when we consider the
regenerated light due to an absorption of TeV photons by CIB, and
estimating its flux is a nontrivial problem; if the regenerated light
significantly changes the high-energy emission profile, it gives a quite
large amount of uncertainty on the GRB physics since the CIB background
as well as extragalactic magnetic field strength, both of which are not
satisfactorily constrained yet, alters the expected signal.
In this paper, therefore, we investigate the effects of the delayed
emission on the primary afterglow light curve in the GeV range; as a
source of the regenerated GeV emission, we consider an afterglow phase
itself, which is well described by an external shock model, while the
other phases have already been investigated in several papers
\citep{Dai02a,Wang04,Razzaque04}.
Among several mechanisms that predict an afterglow spectrum extending to
TeV range, we adopt the IC scattering of the synchrotron photons and
follow the formulation given by \citet{Zhang01} and \citet{Sari01}; this
is because the high-energy emission up to TeV region is most likely
realized due to the IC mechanism \citep{Zhang01}, and the required
values for relevant parameters appear to be realized in many GRBs
\citep{Panaitescu02}.
We show that the regenerated GeV emission, due to the TeV absorption and
following CMB scattering, can only slightly changes the detected
signals.
Therefore, we conclude that the GeV light curves obtained by the future
detectors such as the {\it Gamma-Ray Large Area Space Telescope (GLAST)}
surely give us intrinsic information concerning the GRB fireball, not
affected by the uncertainty of the CIB as well as the extragalactic
magnetic fields.

This paper is organized as follows.
In \S~\ref{sec:High-energy radiation from afterglows}, we briefly
summarize the formulation of the prompt high-energy emission given in
literature, and then in \S~\ref{sec:Interaction with cosmic infrared
background and regenerated high-energy emission}, we describe the
regeneration mechanism from the prompt high-energy emission and give
formulation for that.
The result of the numerical calculation using a reasonable parameter set
is presented in \S~\ref{sec:Results}, and finally, we discuss that
result and give conclusions in \S~\ref{sec:Discussion and conclusions}.

\section{High-energy radiation from afterglows}
\label{sec:High-energy radiation from afterglows}

As for evolution of the fireball and radiation spectrum, we follow the
formulation given in \citet{Zhang01} and \citet{Sari01}, and refer the
reader to the literature for a detailed discussion; here we briefly
summarize necessary information.

The spectrum of the afterglow synchrotron emission \citep*{Sari98} has
breaks at several frequencies, i.e., the self-absorption frequency
$\nu_a$, injection frequency $\nu_m$ corresponding to the minimum
electron Lorentz factor $\gamma_m$, cooling frequency $\nu_c$
corresponding to the electron Lorentz factor $\gamma_c$ for which the
radiative timescale equals the dynamical time, and cutoff frequency
$\nu_u$ corresponding to $\gamma_u$ above which electrons cannot be
accelerated.
Break frequencies relevant for this study and peak flux are given by
\begin{eqnarray}
 \nu_m &=& 2.9\times 10^{16}\,{\rm Hz} ~\epsilon_e^2\epsilon_B^{1/2}
  \mathcal E_{52}^{1/2}t_h^{-3/2}(1+z)^{1/2},\\
 \nu_c &=& 3.1\times 10^{13}\,{\rm Hz}~(1+Y_e)^{-2}\epsilon_B^{-3/2}
  \mathcal E_{52}^{-1/2}n^{-1}\nonumber\\
 &&{}\times t_h^{-1/2}(1+z)^{-1/2},\\
 \nu_u &=& 2.3\times 10^{22}\,{\rm Hz} ~(1+Y_e)^{-1}
  \mathcal E_{52}^{1/8}n^{-1/8}t_h^{-3/8}\nonumber\\
 &&{}\times (1+z)^{-5/8},\\
 F_{\nu,{\rm max}}&=&29\,{\rm mJy}~\epsilon_B^{1/2}
  \mathcal E_{52}n^{1/2}D_{L,28}^{-2}(1+z),
\end{eqnarray}
where $z$ is the redshift of the GRB, $\mathcal E_{52}$ is the fireball
energy per unit solid angle in units of $10^{52}$ ergs sr$^{-1}$, $n$
the external medium density in units of cm$^{-3}$, $\epsilon_e$ and
$\epsilon_B$ represent the fraction of the kinetic energy going to the
electrons and magnetic fields, respectively, $t_h$ is the observer time
measured in hours, and $D_{L,28}$ is the burst luminosity distance
measured in units of $10^{28}$ cm.
The Compton parameter $Y_e$ is given as a ratio between the luminosities
due to IC and synchrotron radiation, and can be represented by $Y_e
=L_{\rm IC}/L_{\rm syn} = [-1+(1+4\eta\epsilon_e/\epsilon_B)^{1/2}]/2$,
where $\eta = \min [1,(\gamma_m/\gamma_c)^{p-2}]$ with $p$ representing
the spectral index of injected electrons \citep{Panaitescu00,Sari01}.

The IC spectrum due to the scattering on the synchrotron seed photons
can be very hard if the Compton parameter $Y_e$ is sufficiently large.
The typical break frequencies that characterize the IC spectrum are
$\nu_m^{\rm IC}\simeq\gamma_m^2\nu_m$ and $\nu_c^{\rm IC}\simeq
\gamma_c^2\nu_c$.
The cutoff frequency in the IC component is defined by $\nu_u^{\rm IC} =
\min (\gamma_u^2\nu_u,\nu_{\rm KN}^{\rm IC})$, where $\nu_{\rm KN}^{\rm
IC}$ is the Klein-Nishina limit, above which the IC cross section is
suppressed.
\citet{Sari01} explicitly gave analytic expressions for the IC spectrum,
and they pointed out that the power-law approximation is no longer
accurate at high-frequency region, on which we focus in this paper.
Therefore, we use their analytic expressions shown in Appendix A of
\citet{Sari01}, on the contrary to \citet{Zhang01}, in which the authors
used an power-law expression for simplicity.
High-energy photons reaching to TeV due to the IC scatterings are
absorbed by the soft photons in the fireball and create the
electron/positron pairs.
We follow the treatment of \citet{Zhang01} for this intrinsic
absorption (see also, \citealt{Lithwick01}; \citealt{Coppi90};
\citealt{Bottcher97}; \citealt{Dermer00b}).

Until this point, we described the emission property in the case that
the density profile of surrounding matter is uniform, i.e., $n$ does not
depend on the radius.
We also consider the case of the wind density profile, which is possibly
the case because the GRB progenitors can eject envelope as a stellar
wind; assuming a constant speed of the wind, the density profile becomes
$n(r) = Ar^{-2}$, where $A$ is a constant independent of radius $r$.
We normalize this constant $A$ as $A=3.0\times 10^{35} A_\ast$ cm$^{-1}$
where $A_\ast = (\dot M /10^{-5}M_\odot ~{\rm yr}^{-1})/ (v/10^3 ~{\rm
km} ~{\rm s}^{-1})$ as in \citet{Chevalier00} for a Wolf-Rayet star.
The relevant frequencies and the peak flux are then be represented by
\begin{eqnarray}
 \nu_m &=& 2.8\times 10^{16}\,{\rm Hz} ~\epsilon_e^2\epsilon_B^{1/2}
  \mathcal E_{52}^{1/2}t_h^{-3/2}(1+z)^{1/2},\\
 \nu_c &=& 2.4\times 10^{11}\,{\rm Hz}~(1+Y_e)^{-2}\epsilon_B^{-3/2}
  \mathcal E_{52}^{1/2}A_\ast^{-2}\nonumber\\
 &&{}\times t_h^{1/2}(1+z)^{-3/2},\\
 \nu_u &=& 1.3\times 10^{22}\,{\rm Hz} ~(1+Y_e)^{-1}
  \mathcal E_{52}^{1/4}A_\ast^{-1/4}t_h^{-1/4}\nonumber\\
 &&{}\times (1+z)^{-3/4},\\
 F_{\nu,{\rm max}}&=&0.33\,{\rm Jy}~\epsilon_B^{1/2}\mathcal E_{52}^{1/2}
  A_\ast t_h^{-1/2} D_{L,28}^{-2}(1+z)^{3/2},
\end{eqnarray}
following the discussion given in \citet{Zhang01}, which is applied to
the case of wind profile.
Both the synchrotron and IC spectra at some fixed time are obtained by
using the same procedure already given above, but the time evolution of
these spectra changes since the dynamics of an expanding jet differs
from the case of constant medium.
Although we do not give full representation of the spectral evolution,
the reader is referred to \citet{Panaitescu00} for analytic treatment
including the wind-like structure.

\section{Interaction with cosmic infrared background and regenerated
 high-energy emission}
\label{sec:Interaction with cosmic infrared background and regenerated
high-energy emission}

For typical GRB locations at redshift $z=1$, \citet{Salamon98} indicated
that the optical depth due to the CIB radiation field reaches $\sim 10$
when the energy of {\it prompt} emission is higher than 300 GeV.
We assume that the electron and positron of the $e^\pm$ pair share 1/2
the photon energy, i.e., $\gamma_e =\epsilon_\gamma /2m_e$.
With this assumption, the created electron/positron spectrum can be
described by
\begin{equation}
 \frac{d^2N_e}{dt_pd\gamma_e}=2\frac{d\nu}{d\gamma_e}
  \frac{F_\nu (t_p,\epsilon_\gamma)}{\epsilon_\gamma}
 =\frac{2}{h\gamma_e}F_\nu (t_p,2m_e\gamma_e),
 \label{eq:electron/positron spectrum}
\end{equation}
where $\epsilon_\gamma =h\nu$, $t_p$ represents the observed time of the
prompt emission provided that there is no absorption by the CIB, and
$F_\nu$ is the flux of prompt photons including the intrinsic
absorption.
Following the result of \citet{Salamon98}, we assume that the
high-energy photons with $\epsilon_\gamma > 300$ GeV are completely
attenuated, creating $e^\pm$ pairs with $\gamma_e > 3\times 10^5$.
The pair creations typically occur at the distance $R_{\rm pair} =
(0.26\sigma_T n_{\rm IR})^{-1} \simeq 5.8\times 10^{24}\,{\rm cm}^{-3}
(n_{\rm IR}/1\,{\rm cm}^{-3})^{-1}$, where $n_{\rm IR}$ is the CIB
number density; this length scale is much less than the distance from
the observer to the GRB, $D_L$, and hence, the attenuation of the
primary photons can be regarded as quite local phenomenon.

The secondary electron/positron pairs then IC scatter the CMB photons up
to GeV energy scale, and pairs cool on a timescale $t_{\rm IC} = 3m_ec/
(4\gamma_e \sigma_Tu_{\rm cmb}(z))\approx 7.3\times 10^{13} (\gamma_e/10^6)
^{-1} (1+z)^{-4}$ s in the local rest frame, where $u_{\rm cmb}(z)$
represents the CMB energy density at redshift $z$.
The IC spectrum from an electron (positron) with the Lorentz factor
$\gamma_e$, scattering on a CMB photon whose energy is $\epsilon_{\rm
cmb}$, $d^3N_{\gamma}/d\epsilon_{\rm cmb}dt_d^\prime dE_\gamma$, is
explicitly given by \citet{Blumenthal70} to be
\begin{eqnarray}
 \frac{d^3N_\gamma}{d\epsilon_{\rm cmb}dt_d^\prime dE_\gamma}
  =\frac{\pi r_0^2c}{2\gamma_e^2}
  \frac{n_{\rm cmb}(\epsilon_{\rm cmb},z)}{\epsilon_{\rm cmb}^2}
     \left(2E_\gamma\ln\frac{E_\gamma}{4\gamma_e^2\epsilon_{\rm cmb}}\right.
  \nonumber\\
 {}\left.
   +E_\gamma+4\gamma_e^2\epsilon_{\rm cmb}
   -\frac{E_\gamma^2}{2\gamma_e^2\epsilon_{\rm cmb}}\right),
 \label{eq:IC spectrum}
\end{eqnarray}
where $n_{\rm cmb}(\epsilon_{\rm cmb},z)$ is the number spectrum of the
CMB at redshift $z$, $t_d^\prime$ represents the time of the delayed
emission in the local rest frame, measured from the onset of the $e^\pm$
pair generation, and $E_\gamma$ the energy of the delayed $\gamma$-ray.
Since the observed time of the delayed emission $t$ can be represented
by $t=t_p+t_d$, where $t_d$ is the observed time of the delayed emission
measured from the pair generation, the flux of the regenerated
$\gamma$-ray is obtained by
\begin{equation}
 F_\nu (t,E_\gamma)=\int_0^t dt_p
  ~E_\gamma \left.\frac{d^3N_{\rm delayed~IC}}{dt_pdt_ddE_\gamma}
	    \right|_{t_d=t-t_p},
  \label{eq:flux of delayed IC}
\end{equation}
where
\begin{eqnarray}
 \frac{d^3N_{\rm delayed~IC}}{dt_pdt_ddE_\gamma}
  =\int d\epsilon_{\rm cmb}\int d\gamma_e
  \left(\frac{d^2N_e}{dt_pd\gamma_e}\right)
  \nonumber\\{}\times
  \left(\frac{d^3N_\gamma}{d\epsilon_{\rm cmb}dt_d^\prime 
   dE_\gamma}\right)t_{\rm IC}(\gamma_e)
  \frac{e^{-t_d/\Delta t(\gamma_e)}}{\Delta t(\gamma_e)},
  \label{eq:integrated function}
\end{eqnarray}
which can be calculated with the previously evaluated spectra (eqs.
[\ref{eq:electron/positron spectrum}] and [\ref{eq:IC spectrum}]).
Here, the lower bound of the integration over $\gamma_e$ is $\max
[3\times 10^5,(E_\gamma /\epsilon_{\rm cmb})^{1/2}/2]$.
In equation (\ref{eq:integrated function}), $(d^3N_\gamma/d\epsilon_{\rm
cmb}dt_d^\prime dE_\gamma) t_{\rm IC}$ shows a total number of the IC
photons per unit CMB energy per unit IC photon energy, emitted until the
parent electron with $\gamma_e$ cools.
The last part of the same equation $e^{-t_d/\Delta t}/\Delta t$
represents the time profile of the delayed $\gamma$-ray emission, and
$\Delta t(\gamma_e)$ is the typical observed duration of the IC photons
from the electron (positron) with Lorentz factor $\gamma_e$.
The typical timescale of the delayed emission measured in the observer
frame is given by $\Delta t(\gamma_e) =\max (\Delta t_{\rm IC}, \Delta
t_A, \Delta t_B)$, where $\Delta t_{\rm IC} = (1+z)t_{\rm
IC}/2\gamma_e^2$ is the IC cooling time; $\Delta t_A = (1+z)R_{\rm
pair}/2\gamma_e^2c$ is the angular spreading time; and $\Delta t_B =
(1+z)t_{\rm IC}\theta_B^2/2$ is the delay time due to magnetic
deflection.
The deflection angle $\theta_B$ is given by $\theta_B \approx 1.3\times
10^{-5} (\gamma_e/10^6)^{-2} B_{{\rm IG},-20}$, where $B_{{\rm IG},
-20}$ represents the extragalactic magnetic field strength in units of
$10^{-20}$ G.

\section{Results}
\label{sec:Results}

We calculated the expected high-energy signal in the GeV range due to
the prompt and regenerated afterglow emissions using equation
(\ref{eq:flux of delayed IC}) as well as the formulation given by
\citet{Zhang01} and \citet{Sari01}.
In the following discussion, we fix several relevant parameters used in
our calculation as follows: $\mathcal E_{52}=10, n=1, A_\ast =1, z = 1$,
and $B_{{\rm IG},-20} = 1$.

\begin{figure}
\begin{center}
\includegraphics[width=8cm]{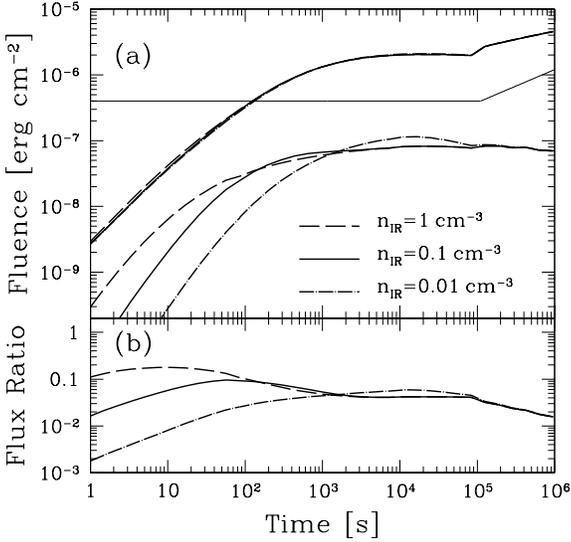}
\caption{(a) Fluence $\int_{\nu_1}^{\nu_2}F_\nu d\nu ~t$ as a function
 of observed time $t$, integrated over 400 MeV to 200 GeV. Lower three
 curves show the fluence of the regenerated emission in the case of
 $n_{\rm IR} = 1$ (dashed curve), 0.1 (solid curve), and 0.01 cm$^{-3}$
 (dot-dashed curve). Upper three (almost degenerate) curves show the
 total fluence. The values for the relevant parameters are: $\mathcal
 E_{52}=10, n=1, \epsilon_e = 0.5, \epsilon_B = 0.01, z = 1$, and
 $B_{{\rm IG},-20} = 1$. The sensitivity curve of the {\it GLAST}
 satellite is also shown. (b) Flux ratio of the delayed and prompt
 afterglow emission, integrated over the same energy range.}
 \label{fig:GeV_light_curve}
\end{center}
\end{figure}

As for the parameters $\epsilon_e$ and $\epsilon_B$, we first fix them
at $0.5$ and $0.01$, respectively.
Figure \ref{fig:GeV_light_curve}(a) shows a fluence $\int_{\nu_1}^
{\nu_2} F_\nu d\nu ~t$ as a function of observed time $t$, where $h\nu_1
= 400$ MeV and $h\nu_2 = 200$ GeV; the lower three curves represent the
fluence of the regenerated emission in the case of $n_{\rm IR} = 1$
(dashed curve), 0.1 (solid curve), and 0.01 cm$^{-3}$ (dot-dashed
curve).
Total fluence from the prompt and delayed afterglow components are shown
as upper three (almost degenerate) curves using the same line type
according to the CIB density.
The fluence threshold for the {\it GLAST} satellite is roughly $\sim
4\times 10^{-7} (t/10^5~{\rm s})^{1/2}$ ergs cm$^{-2}$ for a long
integration time regime (exposure time $t\ga 10^5$ s) and $\sim 4\times
10^{-7}$ ergs cm$^{-2}$ for a short integration time, following the
criterion that at least 5 photons are collected
\citep{Gehrels99,Zhang01}; this sensitivity curve is also plotted in the
same figure as a thin solid line.

Figure \ref{fig:GeV_light_curve}(b) shows a ratio of the regenerated and
prompt flux integrated over the same energy range.
From these figures, it is found that the contribution from the
regenerated GeV emission due to the absorption by the CIB photons peaks
around 10--10$^4$ s after the onset of the afterglow, according to the
CIB density.
This is because the time delay occurs mainly by the angular spreading at
the location of the absorption, $\Delta t_A\propto n_{\rm IR}^{-1}$.
The regenerated emission is expected to only slightly change the
afterglow light curve even when we adopt a rather large value of
$n_{\rm IR}$; for $n_{\rm IR}=1$ cm$^{-3}$, its contribution reaches
$\sim 20$\% of the prompt emission around 10 s, but the total fluence
around that time is far below the detection threshold.

\begin{figure}
\begin{center}
\includegraphics[width=8cm]{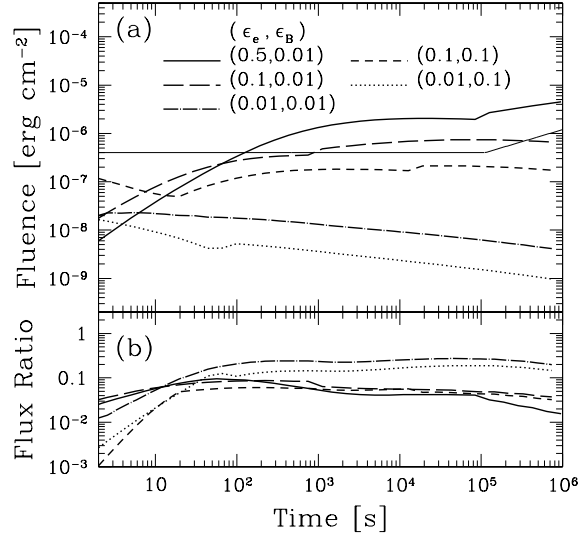}
\caption{The same as figure \ref{fig:GeV_light_curve}, but for various
 sets of $(\epsilon_e,\epsilon_B)$ with fixed value of $n_{\rm IR}=0.1$
 cm$^{-3}$. Total fluence alone is shown in the upper panel (a).}
\label{fig:GeV_light_curve_eB}
\end{center}
\end{figure}

By fixing $n_{\rm IR}$ to be 0.1 cm$^{-3}$, we then investigated the
dependence of the GeV light-curve on $\epsilon_e$ and $\epsilon_B$; the
result is shown in figure \ref{fig:GeV_light_curve_eB}.
Curves in figure \ref{fig:GeV_light_curve_eB}(a) indicate the total
fluence evaluated using various sets of ($\epsilon_e,\epsilon_B$), and
the ratio of regenerated and prompt emission is shown in figure
\ref{fig:GeV_light_curve_eB}(b).
As we expect, rather large values of $\epsilon_e$ are favourable for
possible detection by the {\it GLAST}, because they make the spectrum
extend to high-energy region owing to the IC scattering.
In the case of the small $\epsilon_e$, on the other hand, the flux in
the GeV region is not as strong as the case of large $\epsilon_e$, as
already discussed in several past papers (e.g., \citealt{Zhang01}).
Regardless of the detectability of the total emission, it is easily
found that the regenerated emission is very weak compared with the
prompt one, at most $\sim 30$\% if the value of $\epsilon_B$ is large,
as shown in figure \ref{fig:GeV_light_curve_eB}(b).

\begin{figure}
\begin{center}
\includegraphics[width=8cm]{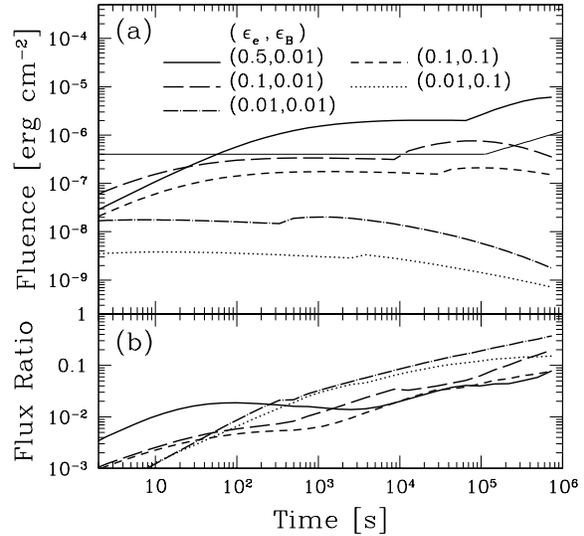}
\caption{The same as figure \ref{fig:GeV_light_curve_eB}, but for the
 wind-like profile of surrounding medium with $A_\ast = 1$.}
\label{fig:GeV_light_curve_wind}
\end{center}
\end{figure}

The result of the same calculation is shown for the case of wind-like
profile of surrounding matter, $n(r)\propto r^{-2}$, in figure
\ref{fig:GeV_light_curve_wind} for various values of ($\epsilon_e,
\epsilon_B$); other parameters are the same as figure
\ref{fig:GeV_light_curve_eB} except for $A_\ast = 1$.
We can confirm that the GeV light-curves as well as their parameter
dependence are basically similar to the case of constant density
profile.
The fraction of the regenerated flux to the prompt one shown in figure
\ref{fig:GeV_light_curve_wind}, on the other hand, behaves somewhat
differently; it increases as the time pasts.
However, it reaches only less than 40\% at $10^6$ s after the onset
of the afterglow, after which the detection threshold for the fluence
glows as $t^{1/2}$ and the detection itself becomes more and more
difficult.
Furthermore, favourable model with large $\epsilon_e$ gives smaller
contribution from the regenerated emission than the models with small
$\epsilon_e$, which is not favoured from the viewpoint of detectability.
In consequence, the regenerated emission gives only very slight
correction to the prompt afterglow emission in the GeV region, and
further, it is found that this characteristic is considerably
independent of the relevant parameters such as $n_{\rm IR}$,
$\epsilon_e$, and $\epsilon_B$ as well as the density profile of the
surrounding medium.

\section{Discussion and conclusions}
\label{sec:Discussion and conclusions}

In recent years, it is suggested that the delayed GeV emission, due to
the absorption of the TeV photons by the CIB radiation field and the
following IC scattering on the CMB photons, may be detected by the
future high-energy detectors such as the {\it GLAST}.
As a source of the original TeV emissions, the internal shocks
\citep{Dai02a,Razzaque04} as well as the initial phase of the external
shocks \citep{Wang04} have been considered.
These authors claim that the delayed emission due to the regeneration
process during propagation can be distinguished from the direct GeV
component due to the IC emission in the afterglow phase.

The emission from the afterglows is also expected to extend to TeV
region with reasonable choices of relevant parameters, and then the
afterglow phase itself can also be a source of the regenerated emission.
Since an estimation of this regenerated emission has not been performed
yet, and further, whether its intensity is above the detection threshold
is nontrivial question, we investigated in this paper the evolution of
the regenerated light from the afterglows, and discussed its
detectability.
As an original high-energy emission model that extends to TeV region, we
have used the modeling of the synchrotron self-IC mechanism by
\citet{Zhang01} and \citet{Sari01}, and also used reasonable choices of
relevant parameters.

As the result of calculation using formalism summarized in
\S~\ref{sec:Interaction with cosmic infrared background and regenerated
high-energy emission} and the constant density profile as well as
$\epsilon_e = 0.5$ and $\epsilon_B = 0.01$, we found that the
contribution of the regeneration of GeV photons could give a correction
at most $\sim 20$\% even if rather large value for the CIB density
($n_{\rm IR}=1$ cm$^{-3}$) is used as shown in figure
\ref{fig:GeV_light_curve}.
Although the CIB density around $z=1$ is unknown, observations suggest
that the local CIB flux at 2.2 $\mu$m is of the order of 10 nW m$^{-2}$
sr$^{-1}$ \citep{Wright01,Wright04}, which corresponds to $n_{\rm IR} =
0.45 \times 10^{-2}$ cm$^{-3}$.
Theoretical model by \citet{Salamon98} indicates that the comoving
density of the CIB photons does not change largely (i.e., less than
factor of $\sim 3$) from $z=1$ to 0, and therefore the proper density of
the CIB might be estimated to be around 0.1 cm$^{-3}$, although there
remains a fair amount of ambiguity.
Our calculation suggests that even if we take a fairly large value for
the CIB density, the regenerated emission cannot change the shape of the
intrinsic light curve, and further, its intensity is far below the
detection threshold of the {\it GLAST} satellite.

In addition to the cosmic CIB density, extragalactic magnetic field
strength may affect the results of our calculation via the value of
$\Delta t_B$ appearing in equation (\ref{eq:integrated function}) if it
is larger than $10^{-20}$ G that we used throughout the above
discussions.
The strength of extragalactic magnetic fields has not been determined
thus far.
Faraday rotation measures imply an upper limit of $\sim 10^{-9}$ G for a
field with 1 Mpc correlation length (see \citealt{Kronberg94} for a
review).
Other methods were proposed to probe fields in the range $10^{-10}$ to
$10^{-21}$ G (\citealt*{Lee95}; \citealt{Plaga95}; \citealt{Guetta03}).
To interpret the observed microgauss magnetic fields in galaxies and
X-ray clusters, the seed fields required in dynamo theories could be as
low as $10^{-20}$ G \citep{Kulsrud97,Kulsrud99}.
Theoretical calculations of primordial magnetic fields show that these
fields could be of order $10^{-20}$ G or even as low as $10^{-29}$ G,
generated during the cosmological QCD or electroweak phase transition,
respectively \citep*{Sigl97}.
Hence, although we used the value of $10^{-20}$ G as the extragalactic
magnetic field strength, it is accompanied by a quite large amount of
uncertainty and it may affect the results give above.
From figure \ref{fig:GeV_light_curve}, it is found that when $B_{\rm IG}
=10^{-20}$ G, the time delay of the regenerated emission is mainly
dominated by angular spreading, i.e., $\Delta t = \Delta t_A$, because
the delayed component changes according to the CIB density $n_{\rm
IR}$.
For further smaller values of $B_{\rm IG}$ than $10^{-20}$ G, therefore,
our conclusion does not change.
On the other hand, if its value is sufficiently large such that the
condition $\Delta t_B > \Delta t_A$ is satisfied for the majority of
possible $\gamma_e$, it further suppresses the regenerated emission
since its flux is inversely proportional to $\Delta t_B$ as clearly
shown in equation (\ref{eq:integrated function}).
In consequence, even if the value of extragalactic magnetic fields
differs from our reference value, our central conclusion that the
regenerated emission is negligibly weak compared with the prompt one
does not change.

We also performed the same calculation but by focusing on dependence on
the relevant parameters ($\epsilon_e,\epsilon_B$) that strongly affect
the high-energy emission mechanism.
We showed that although the total fluence in the GeV region considerably
depends on the values of ($\epsilon_e,\epsilon_B$), but the fraction of
the regenerated emission to the prompt one is always small for any
choices of parameter sets (figure \ref{fig:GeV_light_curve_eB}).
This characteristic holds in the case of the wind-like profile of
surrounding medium as shown in figure \ref{fig:GeV_light_curve_wind}.
All of these facts given above enable us to probe a high-energy emission
mechanism in the GRB fireballs when data of the GeV photons are
accumulated, because the expected signal would be almost completely free
of a large amount of uncertainty concerning the CIB density as well as
extragalactic magnetic field; they never affect the afterglow emission
itself for any choices of the relevant parameters ($\epsilon_e,
\epsilon_B$), whichever (constant or wind-like) profile of the
surrounding medium is truly realized.

\section*{Acknowledgments}

This work was supported by a Grant-in-Aid for JSPS Fellows.

\bibliography{refs}

\label{lastpage}

\end{document}